\documentclass[epj,nopacs]{svjour}
\usepackage{amsmath}
\usepackage{amssymb}
\usepackage{latexsym}
\usepackage[dvips]{graphicx}
\usepackage[english]{babel}

\def\bge{\begin{equation}}
\def\ene{\end{equation}}
\def\bgea{\begin{eqnarray}}
\def\enea{\end{eqnarray}}

\def\bge{\begin{equation}}
\def\ene{\end{equation}}
\def\bgea{\begin{eqnarray}}
\def\enea{\end{eqnarray}}

\def\ls{\raise 1.5pt\hbox{$\,<\;$}\kern -10.5pt\lower3.5pt
          \hbox{$\sim$}\kern 1.5pt} 
\def\gs{\raise 1.5pt\hbox{$\,>\,$}\kern -9.5pt\lower3.5pt
          \hbox{$\sim$}\kern 1.5pt} 

\usepackage{color}

\usepackage{ulem} 

\begin{document}
\sloppy
\title{Cosmological Tests With Strong Gravitational Lenses using Gaussian Processes}
\author{Manoj K. Yennapureddy$^1$ and Fulvio Melia$^2$\thanks{John Woodruff Simpson Fellow.}}
\institute{$^1$Department of Physics, The University of Arizona, AZ 85721, USA,
\email{manojy@email.arizona.edu} \\
$^2$Department of Physics, the Applied Math Program, and Department of Astronomy, \\
              The University of Arizona, Tucson, AZ 85721,
              \email{fmelia@email.arizona.edu}}

\authorrunning{Yennapureddy and Melia}
\titlerunning{Strong Gravitational Lenses}

\date{December 6, 2017}

\abstract{Strong gravitational lenses provide source/lens distance 
ratios $\mathcal{D}_{\rm obs}$ useful in cosmological tests. Previously, 
a catalog of 69 such systems was used in a one-on-one comparison between 
the standard model, $\Lambda$CDM, and the $R_{\rm h}=ct$ universe, which
has thus far been favored by the application of model selection tools
to many other kinds of data. But in that work, the use of model
parametric fits to the observations could not easily distinguish
between these two cosmologies, in part due to the limited measurement
precision. Here, we instead use recently developed methods based on 
Gaussian Processes (GP), in which $\mathcal{D}_{\rm obs}$ may be reconstructed 
directly from the data without assuming any parametric form. This approach
not only smooths out the reconstructed function representing the
data, but also reduces the size of the $1\sigma$ confidence regions,
thereby providing greater power to discern between different models.
With the current sample size, we show that analyzing strong lenses 
with a GP approach can definitely improve the model comparisons, 
producing probability differences in the range $\sim 10-30\%$. These
results are still marginal, however, given the relatively small sample.
Nonetheless, we conclude that the probability of $R_{\rm h}=ct$ being the
correct cosmology is somewhat higher than that of $\Lambda$CDM, with
a degree of significance that grows with the number of sources in the
subsamples we consider. Future surveys will significantly grow the 
catalog of strong lenses and will therefore benefit considerably from the GP 
method we describe here. In addition, we point out that if the $R_{\rm h}=ct$ 
universe is eventually shown to be the correct cosmology, the lack of free 
parameters in the study of strong lenses should provide a remarkably powerful
tool for uncovering the mass structure in lensing galaxies.}
\maketitle

\section{Introduction}
The degree to which light from high-redshift quasars is deflected by
intervening galaxies can be calculated precisely if one has enough
information concerning the distribution of mass within the gravitational
lens \cite{1,2}. Depending on the
mass of the galaxy, and the alignment between source, lens, and observer,
gravitational lenses may be classified either as macro (with sub-classes
of strong and weak lensing) or micro lensing systems. Strong lensing occurs
when the source, lens, and observer are sufficiently well aligned that the
deflection of light forms an Einstein ring. Using the angle of deflection,
one may derive the radius of this ring, from which one may then also
compute the angular diameter distance to the lens. This distance,
however, is model dependent. Hence, together with the measured redshift
of the source, this angular diameter distance may be used to discriminate
between various cosmological models (see, e.g., ref.~\cite{3,4,5,6,7}).

In this paper, we use a recent compilation of 118 \cite{8} plus 40 \cite{9}
strong lensing systems, with good spectroscopic measurements of the central 
velocity dispersion based on the SLACS ({\it Sloan Lens ACS}) Survey \cite{4,10,11}, 
and the LSD ({\it Lenses Structure and Dynamics}) Survey (see, e.g., refs.~\cite{12,13}),
to conduct a comparative study between $\Lambda$CDM \cite{14,15} and
another Friedmann-Robertson-Walker (FRW) cosmology known as the $R_{\rm h}=ct$
universe \cite{16,17}. Over the past decade, such
comparative tests between this alternative model and $\Lambda$CDM have
been carried out using a wide assortment of data, most of them favouring
the former over the latter (for a summary of these tests, see Table~1 in
ref.~\cite{18}). These studies have included high $z$-quasars \cite{19},
gamma-ray bursts \cite{15}, Type Ia SNe \cite{21,22}, and cosmic
chronometers \cite{23}. The $R_{\rm h}=ct$ model is characterized by
a total equation of state $p=-\rho/3$, in terms of the
total pressure $p$ and density $\rho$ in the cosmic fluid.

The results of these comparative tests are not yet universally accepted,
however, and several counterclaims have been made in recent years. One
may loosely group these into four general categories: (1) that the
gravitational radius (and therefore also the Hubble radius) $R_{\rm h}$
is not really physically meaningful \cite{24,25,26}; (2) that the zero active
mass condition $\rho+3p=0$ at the basis of the $R_{\rm h}=ct$ cosmology
is inconsistent with the actual constitutents in the cosmic fluid
\cite{27}; (3) that the $H(z)$ data favour $\Lambda$CDM over
$R_{\rm h}=ct$ \cite{25,28}; and (4) that Type Ia SNe also favour the
concordance model over $R_{\rm h}=ct$ \cite{25,28,29}. These works, and
papers published in response to them \cite{17,23,30,31,32,33},
have generated an important discussion concerning the viability of
$R_{\rm h}=ct$ that we aim to continue here. In \S~7 below, we will
discuss at greater length the need to use truly model-independent data
in these tests, basing their analysis on sound statistical practices.
Such due diligence is of utmost importance in any serious attempt to
compare different cosmologies in an unbiased fashion.

The test most directly relevant to the work reported here was carried out
using strong lenses by ref.~\cite{7}, who based their comparison on
parametric fits from the models themselves, and concluded that both
cosmologies account for the data rather well. The precision of the
measurements used in that application, however, was not good enough to
favour either model over the other. In this paper, we revisit that sample
of strong lensing systems and use an entirely different approach for the
comparison, based on Gaussian Processes (GP) to reconstruct the function representing
the data non-parametrically. In so doing, the angular diameter distance to the
lensing galaxies is determined without pre-assuming any model, providing
a better comparison of the competing cosmologies using a functional area
minimization statistic described in \S~6. An obvious benefit of this approach
is that a reconstructed function representing the data may be found
regardless of whether or not any of the models being tested is actually
the correct cosmology.

In \S~2 of this paper, we describe the lensing equation used in cosmological
tests, and we then describe the data used with this application in \S~3.
The Gaussian processes are summarized in \S~4, and in \S~5 we describe
the cosmological models being tested here. The area minimization statistic
is introduced in \S~6, and we explain how this is used to obtain the
model probabilities. We end with our conclusions in \S~7.

\section{Theory of Lensing}
In work with strong lensing, the observed images are typically fitted using a
singular isothermal ellipsoid approximation (SIE) for the lens \cite{34}.
The projected mass distribution at redshift $z_l$ is assumed
to be elliptical, with semi-major axis $\theta_2$ and semi-minor axis
$\theta_1$. Often, an even simpler approximation suffices, and we make use
of it in this paper: we use a singular isothermal sphere (SIS) for the lens
model, in which the semi-major and semi-minor axes are equal, i.e.,
$\theta_1=\theta_2$. To provide context for this approach, we first
describe SIE lens model and afterwards restrict it further by setting
$\theta_1=\theta_2$. The lens equation \cite{35} that
relates the position $\beta$ in the source plane to the position $\theta$
in the image plane is given by
\begin{equation}
\beta =\theta -\nabla_{\theta} \Phi\;,
\end{equation}
where $\Phi$ is the lensing potential of the SIE given as \cite{36}
\begin{equation}
\Phi=\theta_{\rm E}\sqrt{(1-\epsilon)\,\theta_1^2+(1+\epsilon)\,\theta_2^2}\;,
\end{equation}
and $\epsilon$ is the ellipticity related to the eccentricity according to
\begin{equation}
e=\sqrt{\frac{1-\epsilon}{1+\epsilon}}\;.
\end{equation}
In Equation~(2), $\theta_{\rm E}$ is the Einstein radius, defined as
\begin{equation}
\theta_{\rm E}=4\pi\left(\frac{\sigma_v}{c}\right)^2 \mathcal{D}\;,
\end{equation}
where $\sigma_v$ is the velocity dispersion within the lens and
\begin{equation}
\mathcal{D}\equiv\frac{D_A(z_l,z_s)}{D_A(0,z_s)}\;.
\end{equation}

Notice that Equation~(4) is independent of the Hubble constant $H_0$.
Nonetheless, one must still measure $\sigma_v$, the total velocity
dispersion of stellar and dark matter. Obtaining this quantity is
challenging because it is not the average line-of-sight velocity
dispersion weighted with surface-brightness. The velocity dispersion
of the SIS ($\sigma_{SIS}$ or $\sigma_v)$, may be related to the
central velocity dispersion $\sigma_0$, which is obtained from the
stellar velocity dispersion with one-eighth the effective optical
radius (see, e.g., refs.~\cite{4,5}). Though
this works quite well for massive elliptical galaxies, which
are indistinguishable kinematically from an SIE within one effective
radius, $\sigma_{SIS}$ and $\sigma_0$ are actually not equal. Dark
matter is dynamically hotter than bright stars so the velocity
dispersion of the former must be greater than that of the latter
\cite{37}. Treu et al. \cite{4} studied the homogeneity
of early-type galaxies using the large samples of lenses identified
by the Sloan Lenses ACS Survey (SLACS; \cite{10,38})
and found that $f_{SIS}\equiv\langle {\sigma_{SIS}}/{\sigma_0}\rangle=1.010\pm0.017$
when fitting the geometry of multiple images. Similar results were
found in ref.~\cite{39}, who examined the ratio of stellar
velocity dispersion to $\sigma_{SIS}$ for different anisotropy
parameters. The accumulation of evidence therefore suggests that
$f_{SIS}=1.01$, and this is the value we adopt for
this study. Thus, following ref.~\cite{40}, we write the Einstein
Radius as
\begin{equation}
\theta_{\rm E}=4\pi \left(\frac{\sigma_{SIS}}{c}\right)^2\mathcal{D}\;,
\end{equation}
where
\begin{equation}
\sigma_{SIS}=f_{SIS}\,\sigma_0\;.
\end{equation}

The data based on Equation~(6) will be used to compare our two
cosmological models in this paper. The errors associated with
individual measurements of $\mathcal{D}$ are calculated from the
error propagation equation,
\begin{equation}
\sigma_{\mathcal{D}}=\mathcal{D}_{\rm obs}\left[\left(\frac{\sigma_{\theta_{\rm E}}}
{\theta_{\rm E}}\right)^2+4\left(\frac{\sigma_{\sigma_0}}{\sigma_0}\right)^2+
4\left(\frac{\sigma_f}{f_{SIS}}\right)^2\right]^{1/2}\;,
\end{equation}
containing $\theta_{\rm E}$, $\sigma_0$, $f_{SIS}=1.01$ and $\sigma_f=0.017$.
We follow Grillo et al. \cite{5} and set $\sigma_{\theta_{\rm E}}=0.05\,
\theta_{\rm E}$ and $\sigma_{\sigma_0}=0.05\,\sigma_0$. The overall dispersion in
$\mathcal{D}$ is expected to be $\sigma_{\mathcal{D}}\sim 0.11\mathcal{D}_{\rm obs}$.

\section{Data}
The compilation we use here contains 158 strong lensing systems. These have excellent
spectroscopic measurements of the central velocity dispersion, obtained using the
Sloan Lens ACS (SLACS) Survey \cite{4,10,11} and the LSD (Lenses Structure and Dynamics)
survey \cite{12,13}. One can also find some of the original contributions to these datasets
in refs.~\cite{41,42,43,44,45,46,47,48}. The
velocity dispersion (and its aforementioned error $\sim 5$\%) are obtained from
SDSS (Sloan Digital Sky Survey Database).

\begin{table*}
\center
  \centerline{{\bf Table 1.} Strong Gravitational Lensing Systems with $0.45<z_s<0.475$}\vskip 0.1in
\tiny
  \begin{tabular}{lccccccccc}
                \hline
                \hline
                && \\
                Galaxy&$z_l$&$\theta_{\rm E}$&$\sigma_0$&$\mathcal{D}_{\rm obs}$&$\sigma_\mathcal{D}$&
                $\mathcal{D}_{R_{\rm h}=ct}$&$\mathcal{D}_{\Lambda{\rm CDM}}$&Refs.$^a$ \\
                &&(arc sec)&(km s$^{-1}$)&$f_{SIS}$=1.010
                && \\
                \hline
                && \\
                SDSS J1134 + 6027&0.1528&1.10&$239\pm12$&0.6689&0.735&0.634&0.652&10  \\
                SDSS J1403 + 0006&0.1888&0.83&$213\pm17$&0.635&0.069&0.553&0.573&10 \\
                SDSS J2300 + 0022&0.2285&1.25&$305\pm19$&0.446&0.0512&0.460&0.479&1-9  \\
                SDSS J0956 + 5100&0.2405&1.32&$318\pm17$&0.4536&0.0498&0.441&0.459&1-9  \\
                SDSS J0935 - 0003&0.3475&0.87&$396\pm35$&0.192&0.0211&0.222&0.234&10,11\\
                && \\
                \hline\hline
        \end{tabular}
        \footnotesize \vskip 0.1in
                    $^a$Reference: (1) \cite{50}; (2) \cite{46}; (3) \cite{49};
                    (4) \cite{47}; (5) \cite{48}; (6) \cite{4}; (7) \cite{11}; (8) \cite{5};
                    (9) \cite{6}; (10) \cite{12}; (11) \cite{13}
        \end{table*}

Given that two distances are involved for each lens-source pairing, the GP
method calls for a reconstruction of $\mathcal{D}(z_l,z_s)$ in two dimensions. This
will only be feasible, however, when the sample is large enough to yield enough statistics
to warrant this full approach. For now, even with 158 strong lensing systems, we are
constrained to consider small redshift ranges, effectively reducing the problem to a
one-dimensional reconstruction in each sub-division. Because the data are less dispersed
in the lens plane, where $0.1<z_l<0.5$, and scattered much more in the source plane,
$0.3<z_s<3.0$, we carry out the reconstruction within thin redshift-shells of sources,
turning $D_{\rm obs}(z_l,z_s)$ into a one-dimensional function of $z_l$ for what is essentially
a fixed $z_s$. To minimize the scatter in source redshifts, we use a bin size less than
0.025 and choose those bins that have at least 5 data points within them, allowing us
to reconstruct $D_{\rm obs}(z_l,z_s)$ using GP for each of the selected bins. In our sample
of 158 strong-lensing systems, these criteria therefore allow us to assemble 5 different
redshift bins, with anywhere from 5 to 9 lens-source pairs in each of them. These strong
lenses are displayed in Tables~1 to 5. Note that for the purpose of GP reconstruction
in one dimension, we assume that all the sources in redshift bin $(z_s,z_s+\Delta z)$ have
the same average redshift $z_s+\Delta z/2$.

\begin{table*}
\center
  \centerline{{\bf Table 2.} Strong Gravitational Lensing Systems with $0.46<z_s<0.485$}\vskip 0.1in
\tiny
  \begin{tabular}{lccccccccc}
                \hline
                \hline
                && \\
                Galaxy&$z_l$&$\theta_{\rm E}$&$\sigma_0$&$\mathcal{D}_{\rm obs}$&$\sigma_\mathcal{D}$&
                $\mathcal{D}_{R_{\rm h}=ct}$&$\mathcal{D}_{\Lambda{\rm CDM}}$&Refs.$^b$ \\
                &&(arc sec)&(km s$^{-1}$)&$f_{SIS}$=1.010
                && \\
                \hline
                && \\
                SDSS J1134 + 6027&0.1528&1.10&$239\pm12$&0.6689&0.735&0.634&0.652&10  \\
                SDSS J1403 + 0006&0.1888&0.83&$213\pm17$&0.635&0.069&0.553&0.573&10 \\
                SDSS J1402 + 6321&0.2046&1.39&$290\pm16$&0.5743&0.063&0.526&0.546&1-9 \\
                SDSS J1205 + 4910&0.2150&1.22&$281\pm14$&0.5368&0.059&0.504&0.524&10  \\
                SDSS J2300 + 0022&0.2285&1.25&$305\pm19$&0.446&0.0512&0.460&0.479&1-9  \\
                SDSS J0956 + 5100&0.2405&1.32&$318\pm17$&0.4536&0.0498&0.441&0.459&1-9  \\
                SDSS J0935 - 0003&0.3475&0.87&$396\pm35$&0.192&0.0211&0.222&0.234&10,11\\
                && \\
                \hline\hline
        \end{tabular}
        \footnotesize \vskip 0.1in
                    $^a$Reference: (1) \cite{50}; (2) \cite{46}; (3) \cite{49};
                    (4) \cite{47}; (5) \cite{48}; (6) \cite{4}; (7) \cite{11}; (8) \cite{5};
                    (9) \cite{6}; (10) \cite{12}; (11) \cite{13}
        \end{table*}

\begin{table*}
\center
  \centerline{{\bf Table 3.} Strong Gravitational Lensing Systems with $0.5<z_s<0.525$}\vskip 0.1in
\tiny
  \begin{tabular}{lccccccccc}
                \hline
                \hline
                && \\
                Galaxy&$z_l$&$\theta_{\rm E}$&$\sigma_0$&$\mathcal{D}_{\rm obs}$&$\sigma_\mathcal{D}$&
                $\mathcal{D}_{R_{\rm h}=ct}$&$\mathcal{D}_{\Lambda{\rm CDM}}$&Refs.$^c$ \\
                &&(arc sec)&(km s$^{-1}$)&$f_{SIS}$=1.010
                && \\
                \hline
                && \\
                SDSS J1451 - 0239&0.1254&1.04&$223\pm14$&0.726&0.079&0.718&0.735&10,11  \\
                SDSS J2303 + 1422&0.1553&1.64&$271\pm16$&0.775&0.0852&0.654&0.673&1-9  \\
                SDSS J1627 - 0053&0.2076&1.21&$295\pm13$&0.482&0.052&0.552&0.573&1-9  \\
                SDSS J1142 + 1001&0.2218&0.98&$221\pm22$&0.697&0.0766&0.509&0.529&10,11  \\
                SDSS J0109 + 1500&0.2939&0.69&$251\pm19$&0.3807&0.0418&0.389&0.407&10  \\
                SDSS J0216 - 0813&0.3317&1.15&$349\pm24$&0.3287&0.0361&0.320&0.336&1-9  \\
                && \\
                \hline\hline
        \end{tabular}
        \footnotesize \vskip 0.1in
                    $^a$Reference: (1) \cite{50}; (2) \cite{46}; (3) \cite{49};
                    (4) \cite{47}; (5) \cite{48}; (6) \cite{4}; (7) \cite{11}; (8) \cite{5};
                    (9) \cite{6}; (10) \cite{12}; (11) \cite{13}
         \end{table*}

\section{Gaussian Processes and Model Comparisons}
Adapting the code developed by Seikel et al. ref.~\cite{51} for Gaussian Processes in python, we
reconstruct $\mathcal{D}_{\rm obs}(z_l,\langle z_s\rangle)$ for each of the sub-samples in
Tables~1 to 5, without assuming any model a priori. The GP method uses some of the attributes of a
Gaussian distribution, though the former utilizes a distribution over functions obtained
using GP, while the latter represents a random variable. The reconstruction of a function
$f(x)$ at $x$ using GP creates a Gaussian random variable with mean $\mu(x)$ and variance
$\sigma(x)$. The function reconstructed at $x$ using GP, however, is not independent of
that reconstructed at $\tilde{x}=(x+dx)$, these being related by a covariance function
$k(x,\tilde{x})$. Although one can use many possible forms of $k$, we use one that
depends on the distance between $x$ and $\tilde{x}$, i.e., the squared exponential
covariance function defined as
\begin{equation}
k(x,\tilde{x})=\sigma_f^2\exp\left(-\frac{(x-\tilde{x})^2}{2\Delta^2}\right)\;.
\end{equation}
Note that this function depends on two hyperparameters, $\sigma_f$ and $\Delta$,
where $\sigma_f$ indicates a change in the $y$-direction and $\Delta$ represents a
distance over which a significant change in the $x$-direction occurs. Overall, these 
two hyperparameters characterize the smoothness of the function $k$, and are trained
on the data using a maximum likelihood procedure, which leads to the reconstructed
$\mathcal{D}_{\rm obs}(z_l,\langle z_s\rangle)$ function for each source redshift
shell centered on $z_s$. For this paper, we have found that these hyperparameter
values are $0.144$ and $0.661$, respectively.

\begin{table*}
\center
  \centerline{{\bf Table 4.} Strong Gravitational Lensing Systems with $0.51<z_s<0.535$}\vskip 0.1in
\tiny
  \begin{tabular}{lccccccccc}
                \hline
                \hline
                && \\
                Galaxy&$z_l$&$\theta_{\rm E}$&$\sigma_0$&$\mathcal{D}_{\rm obs}$&$\sigma_\mathcal{D}$&
                $\mathcal{D}_{R_{\rm h}=ct}$&$\mathcal{D}_{\Lambda{\rm CDM}}$&Refs.$^d$ \\
                &&(arc sec)&(km s$^{-1}$)&$f_{SIS}$=1.010
                && \\
                \hline
                && \\
                SDSS J2321 - 0939&0.0819&1.57&$245\pm70$&0.9082&0.0999&0.816&0.829&1-9  \\
                SDSS J1451 - 0239&0.1254&1.04&$223\pm14$&0.726&0.079&0.718&0.735&10,11  \\
                SDSS J0959 + 0410 &0.1260&1.00&$229\pm13$&0.6616&0.07277&0.723&0.740&1-9  \\
                SDSS J1538 - 5817&0.1428&1.00&$189\pm12$&0.9717&0.106&0.687&0.705&10,11  \\
                SDSS J2303 + 1422&0.1553&1.64&$271\pm16$&0.775&0.0852&0.654&0.673&1-9  \\
                SDSS J1627 - 0053&0.2076&1.21&$295\pm13$&0.482&0.052&0.552&0.573&1-9  \\
                SDSS J0959 + 4416&0.2369&0.96&$244\pm19$&0.5597&0.0615&0.501&0.521&10  \\
                SDSS J0109 + 1500&0.2939&0.69&$251\pm19$&0.3807&0.0418&0.389&0.407&10  \\
                SDSS J0216 - 0813&0.3317&1.15&$349\pm24$&0.3287&0.0361&0.320&0.336&1-9  \\
                && \\
                \hline\hline
        \end{tabular}
        \footnotesize \vskip 0.1in
                    $^a$Reference: (1) \cite{50}; (2) \cite{46}; (3) \cite{49};
                    (4) \cite{47}; (5) \cite{48}; (6) \cite{4}; (7) \cite{11}; (8) \cite{5};
                    (9) \cite{6}; (10) \cite{12}; (11) \cite{13}
         \end{table*}

\begin{table*}
\center
  \centerline{{\bf Table 5.} Strong Gravitational Lensing Systems with $0.52<z_s<0.545$}\vskip 0.1in
\tiny
  \begin{tabular}{lccccccccc}
                \hline
                \hline
                && \\
                Galaxy&$z_l$&$\theta_{\rm E}$&$\sigma_0$&$\mathcal{D}_{\rm obs}$&$\sigma_\mathcal{D}$&
                $\mathcal{D}_{R_{\rm h}=ct}$&$\mathcal{D}_{\Lambda{\rm CDM}}$&Refs.$^e$ \\
                &&(arc sec)&(km s$^{-1}$)&$f_{SIS}$=1.010
                && \\
                \hline
                && \\
                SDSS J1420 + 6019&0.0629&1.04&$206\pm5$&0.851&0.0936&0.858&0.869&1-9  \\
                SDSS J2321 - 0939&0.0819&1.57&$245\pm70$&0.9082&0.0999&0.816&0.829&1-9  \\
                SDSS J1451 - 0239&0.1254&1.04&$223\pm14$&0.726&0.079&0.718&0.735&10,11  \\
                SDSS J0959 + 0410 &0.1260&1.00&$229\pm13$&0.6616&0.07277&0.723&0.740&1-9  \\
                SDSS J1538 - 5817&0.1428&1.00&$189\pm12$&0.9717&0.106&0.687&0.705&10,11  \\
                SDSS J1627 - 0053&0.2076&1.21&$295\pm13$&0.482&0.052&0.552&0.573&1-9  \\
                SDSS J0959 + 4416&0.2369&0.96&$244\pm19$&0.5597&0.0615&0.501&0.521&10  \\
                SDSS J0109 + 1500&0.2939&0.69&$251\pm19$&0.3807&0.0418&0.389&0.407&10  \\
                SDSS J0216 - 0813&0.3317&1.15&$349\pm24$&0.3287&0.0361&0.320&0.336&1-9  \\
                && \\
                \hline\hline
        \end{tabular}
        \footnotesize \vskip 0.1in
                    $^a$Reference: (1) \cite{50}; (2) \cite{46}; (3) \cite{49};
                    (4) \cite{47}; (5) \cite{48}; (6) \cite{4}; (7) \cite{11}; (8) \cite{5};
                    (9) \cite{6}; (10) \cite{12}; (11) \cite{13}
         \end{table*}

One of the principal features of the GP approach that we highlight in this application
to strong lenses concerns the estimation of the $1\sigma$ confidence region attached
to the reconstructed $\mathcal{D}_{\rm obs}(z_l,\langle z_s\rangle)$ curves. The
$1\sigma$ confidence region depends on both the actual errors of individual data
points, $\sigma_{\mathcal{D}i}$, on the optimized hyperparameter $\sigma_f$
(see Eq.~9) and on the product $K_*K^{-1}K_*^T$ (see ref.~\cite{51}), where
$K_*$ is the covariance matrix at the point of estimation $x_*$, calculated
using the given data at $x_i$, according to
\begin{equation}
K_*=[k(x_1,x_*),k(x_2,x_*),...,k(x_i,x_*)]\;.
\end{equation}
$K$ is the covariance matrix for the original dataset. Note that the dispersion at point
$x_i$ will be less than $\sigma_{\mathcal{D}i}$ when $K_*K^{-1}K_*^T>\sigma_f$, i.e.,
when for that point of estimation there is a large correlation between the data. From
Equation~(9) it is clear that the correlation between any two points $x$ and $\tilde{x}$
will be large only when $x-\tilde{x}<\sqrt{2}\Delta$. This condition, however, is
satisfied most frequently for the strong lenses used in our study, which results
in GP estimated $1\sigma$ confidence regions that are smaller than the errors in
the original data. We refer the reader to ref.~\cite{51} for further details.

The principal goal of this paper is to use a GP reconstruction of the
$\mathcal{D}_{\rm obs}(z_l,\langle z_s\rangle)$ functions in order to
compare the predictions of the $\Lambda$CDM and $R_{\rm h}=ct$
cosmological models. The standard model contains radiation (photons and
neutrinos), matter (baryonic and dark) and dark energy in the form of
a cosmological constant. This blend of constituents, currently dominated
by dark energy, is producing a phase of accelerated expansion, following
an earlier period of deceleration when radiation was dominant. In terms
of today's critical density $\rho_c\equiv 3c^2H_0/8\pi G$ and Hubble
constant $H_0$, the Hubble expansion rate in this cosmology depends on
the matter density, $\Omega_{\rm m}\equiv{\rho_{\rm m}}/{\rho_c}$,
radiation density, $\Omega_{\rm r}\equiv{\rho_{\rm r}}/{\rho_c}$ and
dark energy density, $\Omega_{\rm de}\equiv{\rho_{\rm de}}/{\rho_c}$,
with the constraint $\Omega_{\rm m}+\Omega_{\rm r}+\Omega_{\rm de}=1$.
Since $\Omega_{\rm r}$ is negligible in the current era, we ignore
radiation and use $\Omega_{\rm de}=1-\Omega_{\rm m}$. For all the
calculations, we use the parameters optimized by {\it Planck},
with $\Omega_{\rm m}=0.272$, and $\Omega_{\rm de}=0.728$. Thus,
one deduces from the Friedmann equation that in $\Lambda$CDM
\begin{equation}
H(z)=H_0 \sqrt{\Omega_{\rm m}(1+z)^3+\Omega_{\rm de}}\;.
\end{equation}
The angular diameter distance between redshifts $z_1$ and $z_2$ is given as
\begin{equation}
D(z_1,z_2)=\frac{1}{1+z_2}\int_{z_1}^{z_2}dz\, \frac{c}{H(z)}\;.
\end{equation}
Therefore, substituting for $H(z)$ from Equation~(11), one gets
\begin{eqnarray}
D_A^{\Lambda{\rm CDM}}(z_1,z_2)&=&\frac{c}{H_0}\frac{1}{(1+z_2)}\times\nonumber\\
\null&\null&\hskip-0.3in\int_{z_1}^{z_2}\left[\Omega_{\rm m}(1+z)^3+\Omega_{\rm de}\right]^{-1/2}dz\;.
\end{eqnarray}
\par

\begin{figure}
\begin{center}
\includegraphics[width=1.0\linewidth]{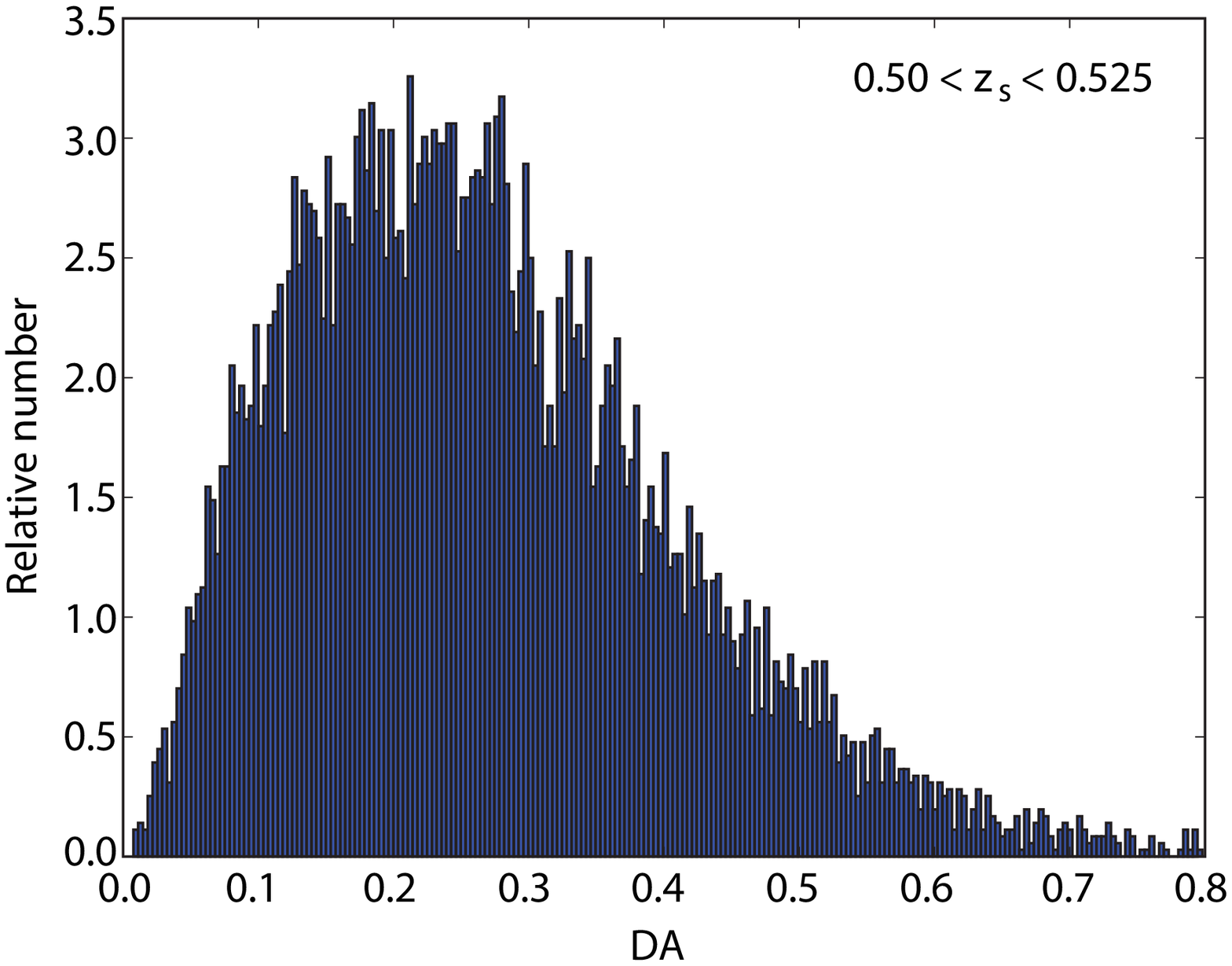}\\
\vskip 0.2in
\includegraphics[width=1.0\linewidth]{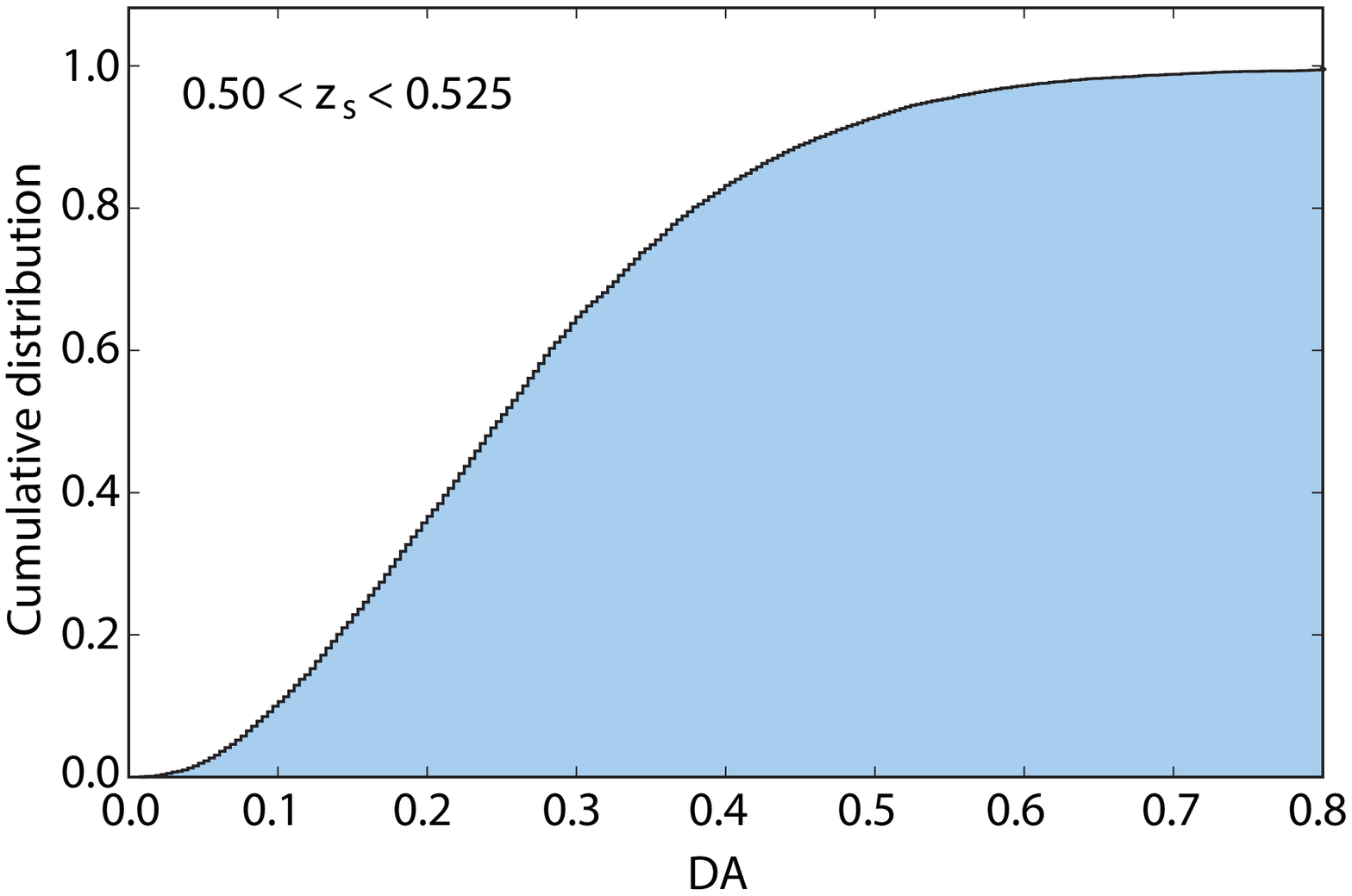}
\end{center}
\caption{{\it Top panel:} The distribution of frequency versus area differential
$DA$ for a mock sample with source shell $0.5<z_s<0.525$; and {\it Bottom panel:}
its corresponding cumulative probability distribution.}
\end{figure}

The $R_{\rm h}=ct$ universe \cite{16,17,52,53,54} is also an FRW cosmology
with radiation (photons and neutrinos), matter (baryonic and dark) and dark
energy, with radiation and dark energy dominating the early Universe, and
matter and dark energy dominating the current era \cite{55}. But while it
is similar to $\Lambda$CDM
in this regard, it has an additional constraint on the total equation of
state, i.e., $\rho+3p=0$, the so-called zero active mass condition, where
$\rho$ and $p$ are the total energy density and pressure, respectively.
With this additional constraint, the $R_{\rm h}=ct$ universe always expands
at a constant rate, which depends on only one parameter---the Hubble constant
$H_0$. Using the Friedmann equation with zero active mass, we find that
\begin{equation}
H^{R_{\rm h}=ct}(z)=H_0(1+z)\;,
\end{equation}
and from Equation~(12), we therefore find that
\begin{equation}
D_A^{R_{\rm h}=ct}(z_1,z_2)=\frac{c}{H_0}\frac{1}{(1+z_2)}\ln\left(\frac{1+z_2}{1+z_1}\right)\;.
\end{equation}

\section{The Area Minimization Statistic}
Now that we are dealing with a comparison between two continuous functions,
i.e., $\mathcal{D}_{\rm obs}$ with either $\mathcal{D}^{\rm \Lambda{\rm CDM}}$ or 
$\mathcal{D}^{R_{\rm h}=ct}$ (each derived from Eq.~5 using Eqs.~13 and 15), we 
cannot use discrete sampling statistics, such as weighted least squares, for the 
comparison of different models. The reason is that sampling at random points to 
obtain the squares of differences between model and reconstructed curve would lose 
information between these points, whose importance cannot be ascertained prior to
the sampling. To overcome this deficiency, we introduce a new statistic, based on
a previous application \cite{56,57}, which we call the ``Area Minimization Statistic" 
to estimate each model's probability of being consistent with the data. Our
principal assumption is that the measurement errors are Gaussian, which we use
to generate a mock sample of GP reconstructed curves representing the possible
variation of $\mathcal{D}$ away from $\mathcal{D}_{\rm obs}$. We do this by
employing the Gaussian randomization
\begin{equation}
\mathcal{D}_i(z_l,\langle z_s\rangle)=\mathcal{D}_{i,\,{\rm obs}}(z_l,\langle z_s\rangle)+r 
\sigma_{\mathcal{D}_i}\;,
\end{equation}
where $\mathcal{D}_{i,\,{\rm obs}}(z_l,\langle z_s\rangle)$ are the actual
measurements as a function of $z_l$ for each source shell $\langle z_s\rangle$.
$\sigma_{\mathcal{D}_i}$ are the actual observed errors and $r$ is a Gaussian
random variable with zero mean and a variance of $1$. Next, these
$\mathcal{D}_i(z_l,\langle z_s\rangle)$ are used together with the errors
$\sigma_{\mathcal{D}_i}$ to reconstruct the function
$\mathcal{D}_{\rm mock}(z_l,\langle z_s\rangle)$
corresponding to each mock sample, and finally we calculate the weighted
absolute area difference between $\mathcal{D}_{\rm mock}(z_l,\langle z_s\rangle)$
and the GP reconstructed function of the actual data according to
\begin{equation}
DA= \int_{z_{\rm min}}^{z_{\rm max}}dz_l\bigg(\frac{\big|\mathcal{D}_{\rm mock}(z_l,\langle z_s\rangle)-
		\mathcal{D}_{\rm obs}(z_l,\langle z_s\rangle)\big|}{\sigma(z_l)}\bigg)\;.
\end{equation}
In this expression, $z_{\rm min}$ and $z_{\rm max}$ are the minimum and maximum
redshifts, respectively, of the data range. We repeat this procedure $10,000$ times 
to build a distribution of frequency versus area differential $DA$, and from it construct 
the cumulative probability distribution. In figure~1 we show these quantities for the
illustrative source shell $0.50<z_s<0.525$ (the frequency is shown in the top panel,
and the cumulative probability distribution is on the bottom). 
This procedure generates a 1-to-1 mapping between the value of $DA$ and the frequency
with which it arises. With the additional assumption that curves with a smaller $DA$
are a better match to $\mathcal{D}_{\rm obs}$, one can then use the cumulative distribution
to estimate the probability that the difference between a model's prediction and the
reconstructed curve is merely due to Gaussian randomness. When comparing a model's
prediction to the data, we therefore calculate its $DA$ and use our 1-to-1 mapping to
determine the probability that its inconsistency with the data is just due to variance, 
rather than the model being wrong. These are the probabilities we then compare to 
determine which model is more likely to be correct. This basic concept is common to 
many kinds of statistical approaches, though none of the existing ones can be used 
when comparing two continuous curves, as we have here.

The reconstructed curves for our 5 subsamples are shown in the left-hand panels 
of figure~2. These correspond to the 5 source redshift shells in Tables~1 to 5. 
The corresponding cumulative probability distributions are plotted in the right-hand 
panels, which also locate the $DA$ values for $R_{\rm h}=ct$ (yellow) and $\Lambda$CDM 
(red). The probabilities associated with these differential areas are summarized in 
Table~6. Along with the reconstructed functions, the left-hand panels also show
the corresponding $1\sigma$ (dark) and $2\sigma$ (light) confidence regions
provided by the GP, and the theoretical predictions in $\Lambda$CDM (dashed)
and $R_{\rm h}=ct$ (dotted). As we highlighted earlier, the functions
$\mathcal{D}_{\rm obs}(z_l,\langle z_s\rangle)$ have been reconstructed
without pre-assuming any parametric form, so in principle they represent
the actual variation of $\mathcal{D}$ with redshift, regardless of whether
or not either of the two models being tested here is the correct cosmology.

The overall impression one gets from the results displayed in figure~2
and summarized in Table~6 is that, for every source redshift shell sampled
here, the probability of $R_{\rm h}=ct$ being consistent with the GP
reconstructed function $\mathcal{D}_{\rm obs}$ is $\sim 10-30\%$ higher than 
that for $\Lambda$CDM. Future surveys will greatly grow the sample of sources
available for this type of analysis, differentiating between these two models
with greater confidence.

\begin{figure}
	\begin{center}
	\begin{tabular}{cc}
\includegraphics[width=0.47\linewidth]{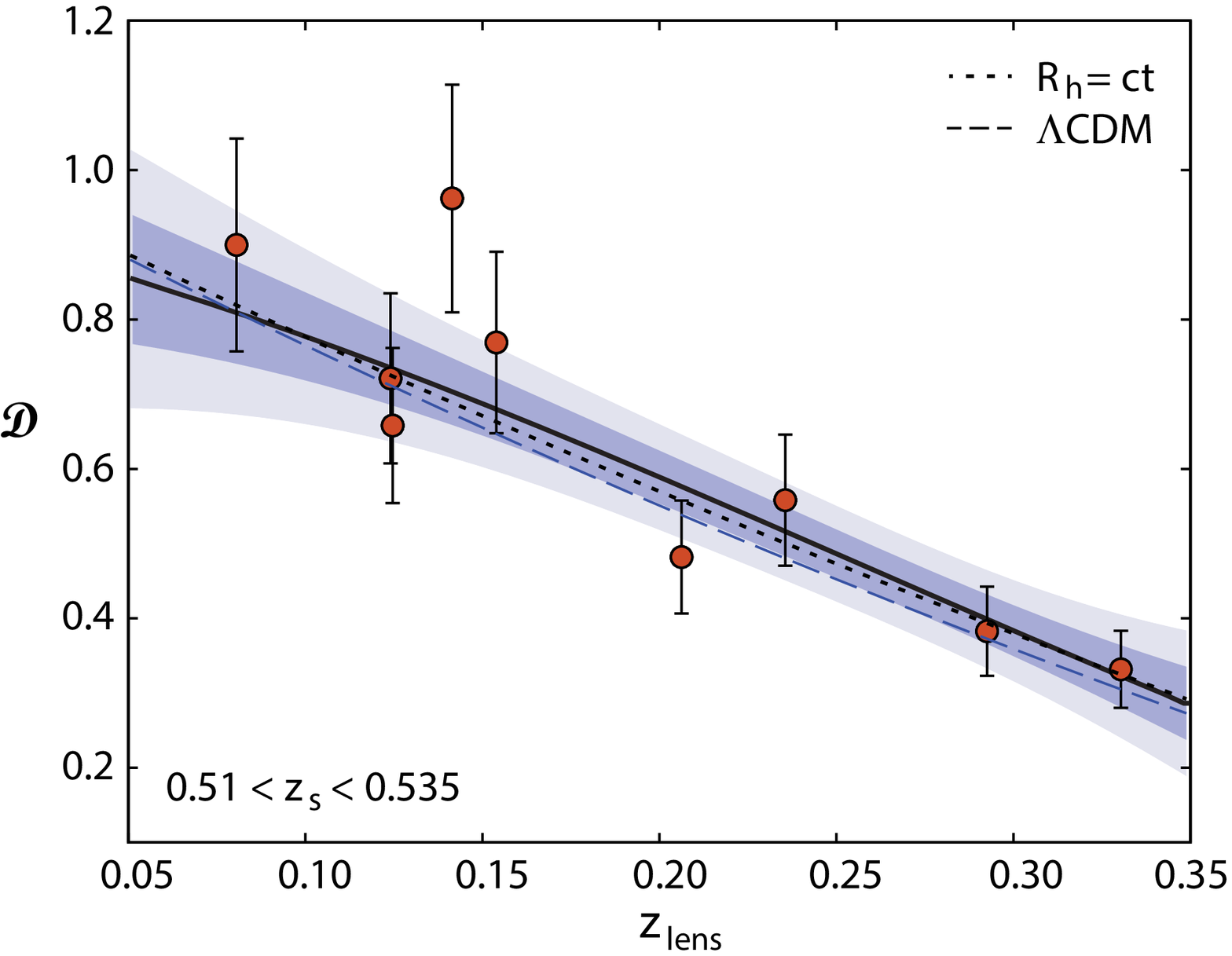}\hskip 0.1in 
\includegraphics[width=0.47\linewidth]{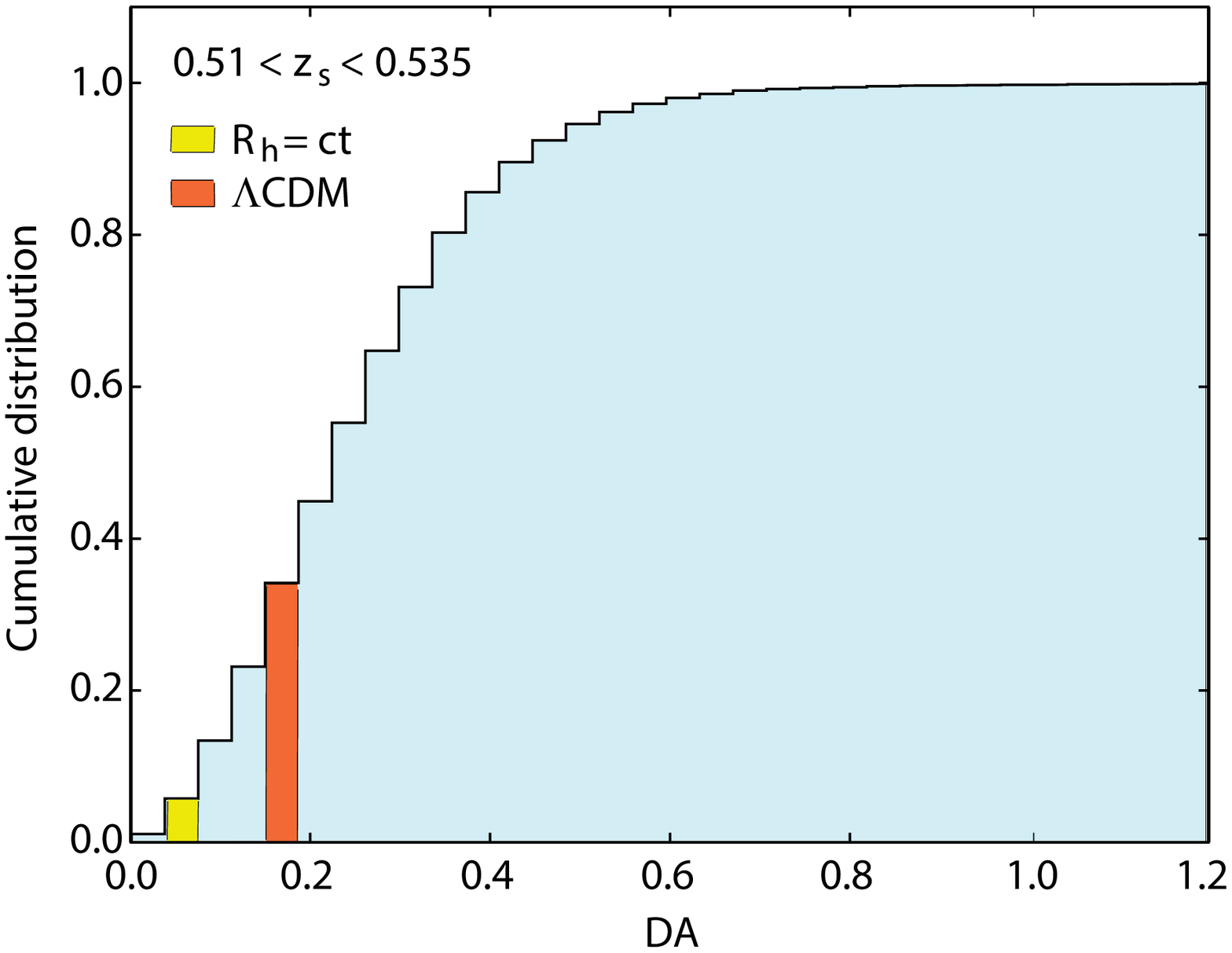} 
\\
\includegraphics[width=0.47\linewidth]{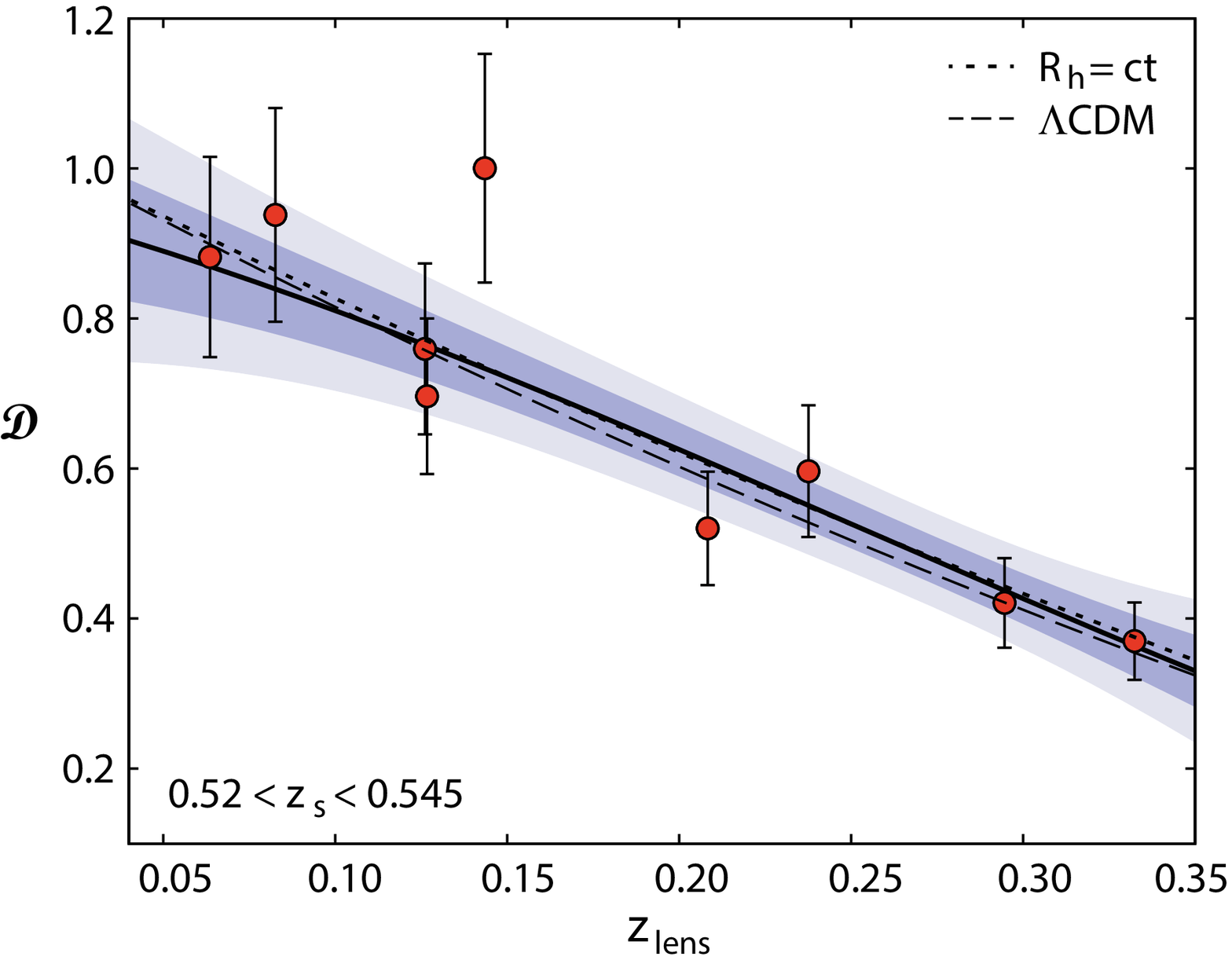}\hskip 0.1in  
\includegraphics[width=0.47\linewidth]{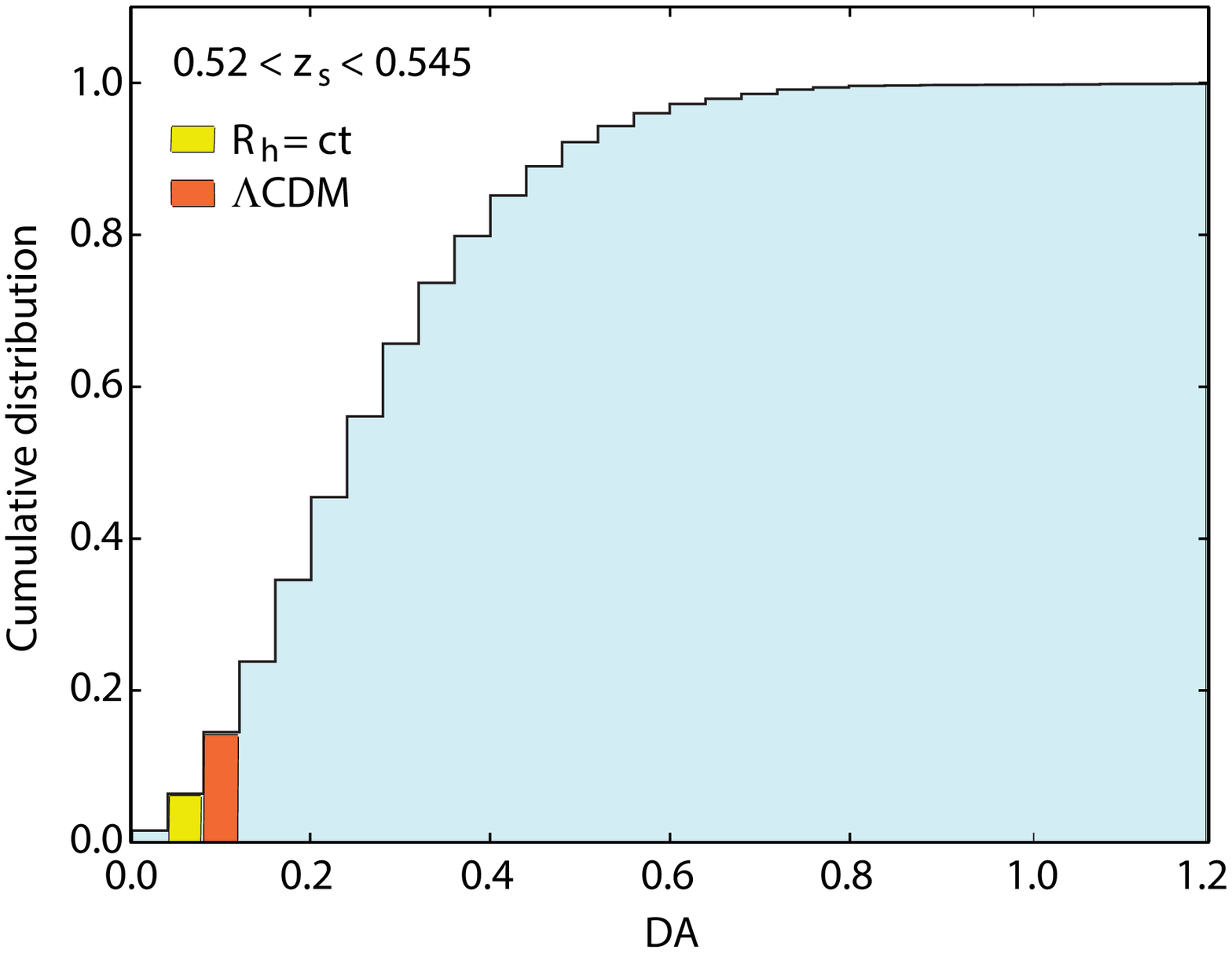} 
\\
\includegraphics[width=0.47\linewidth]{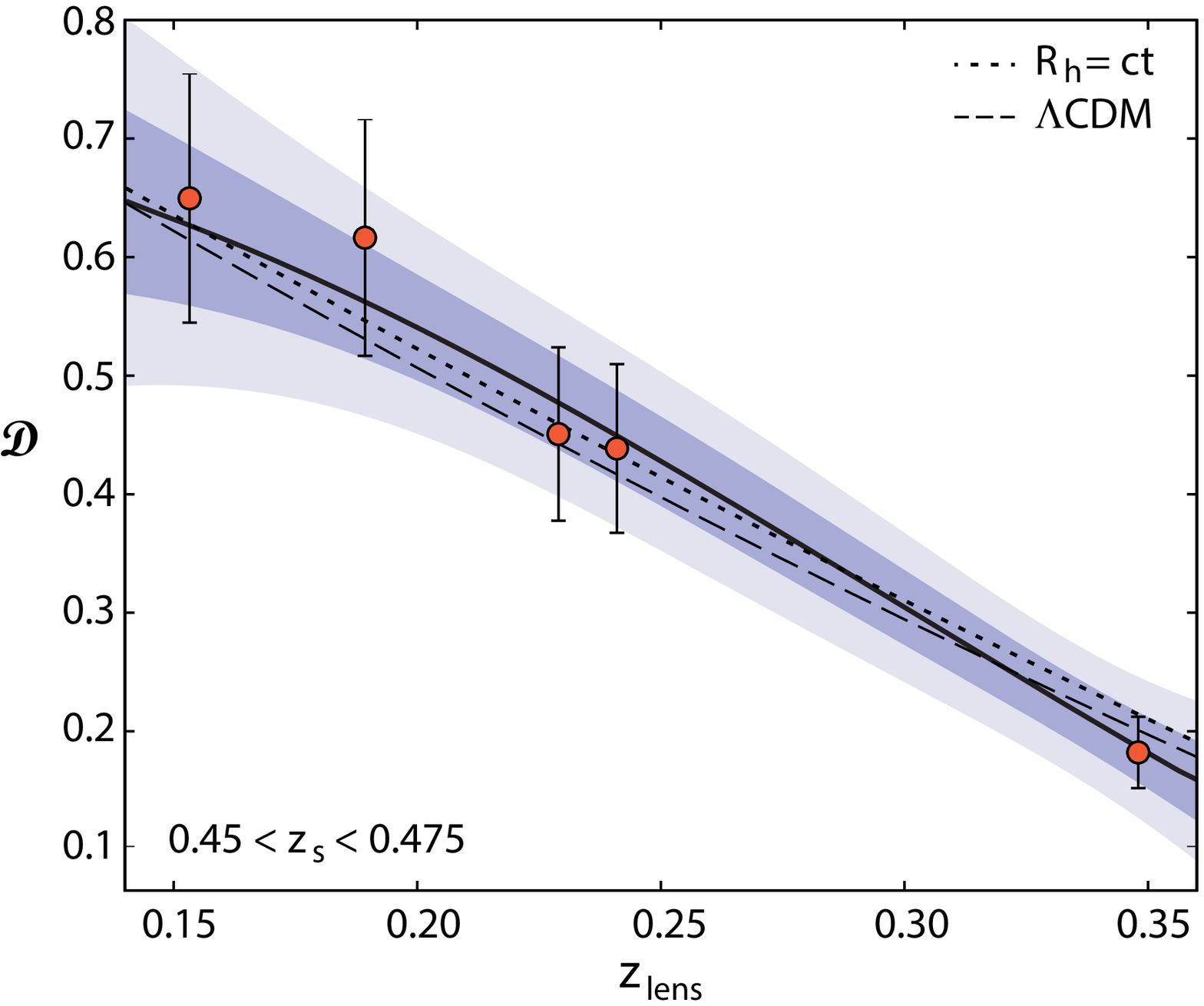}\hskip 0.1in 
\includegraphics[width=0.47\linewidth]{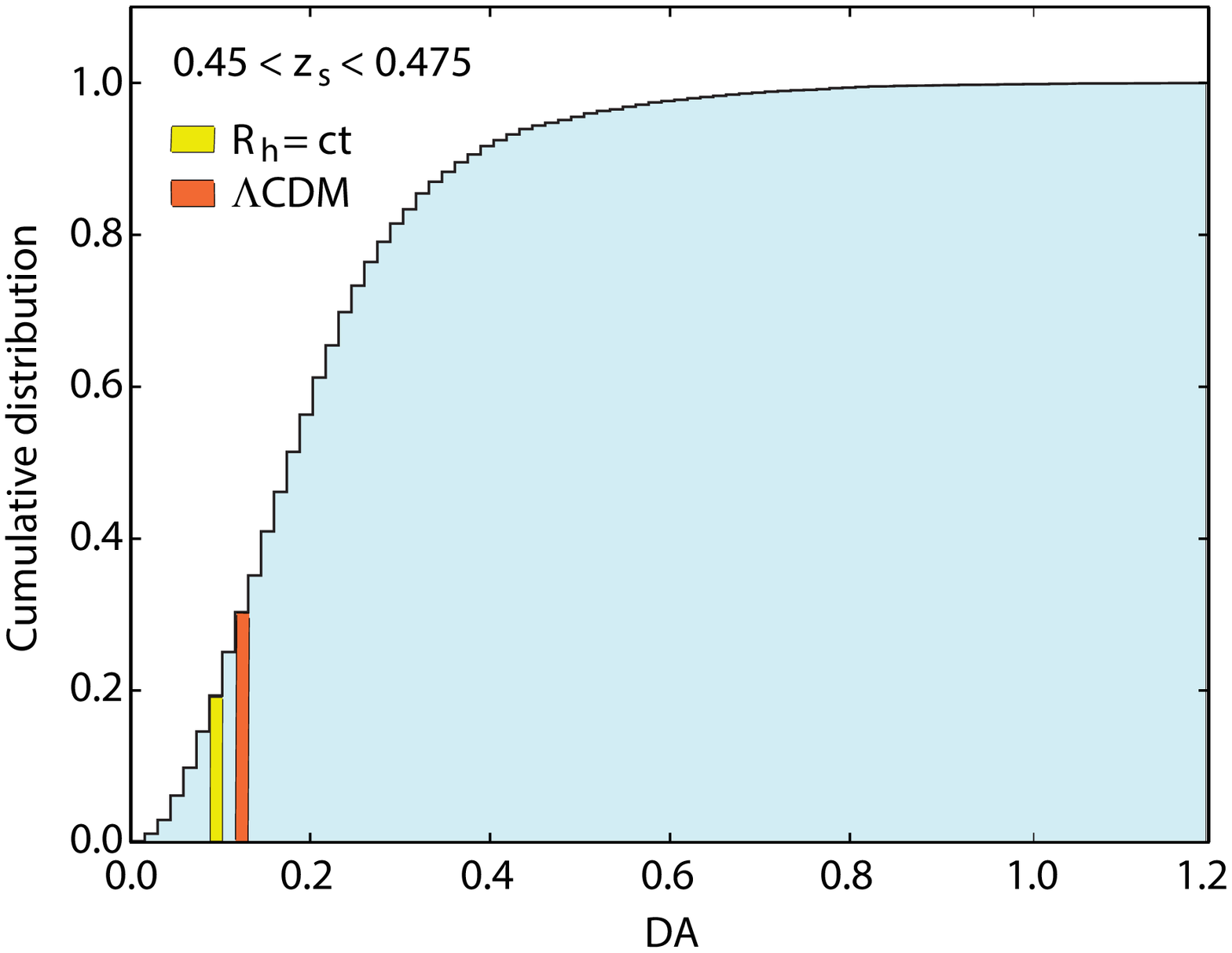} 
\\
\includegraphics[width=0.47\linewidth]{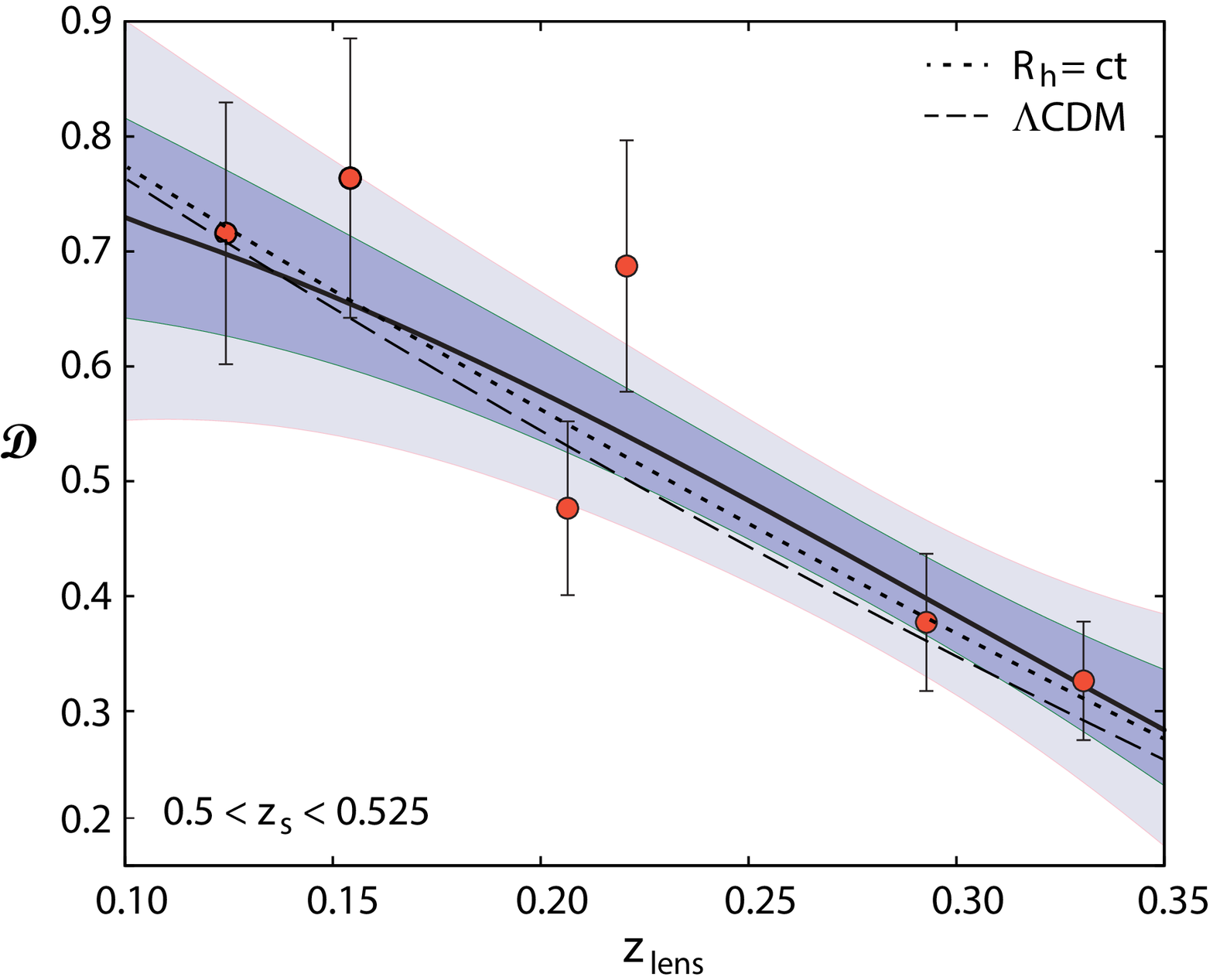}\hskip 0.1in 
\includegraphics[width=0.47\linewidth]{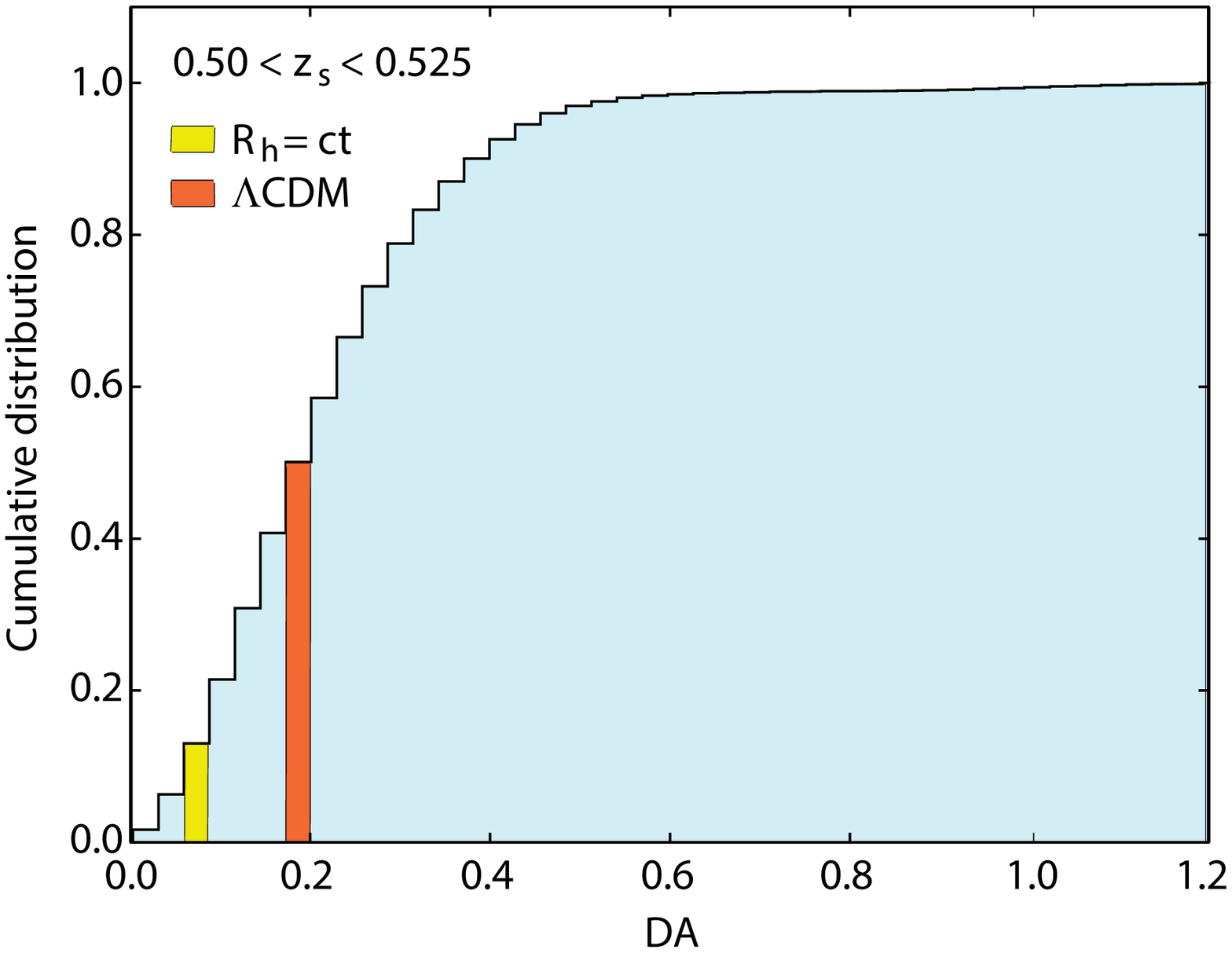}
\\
\includegraphics[width=0.47\linewidth]{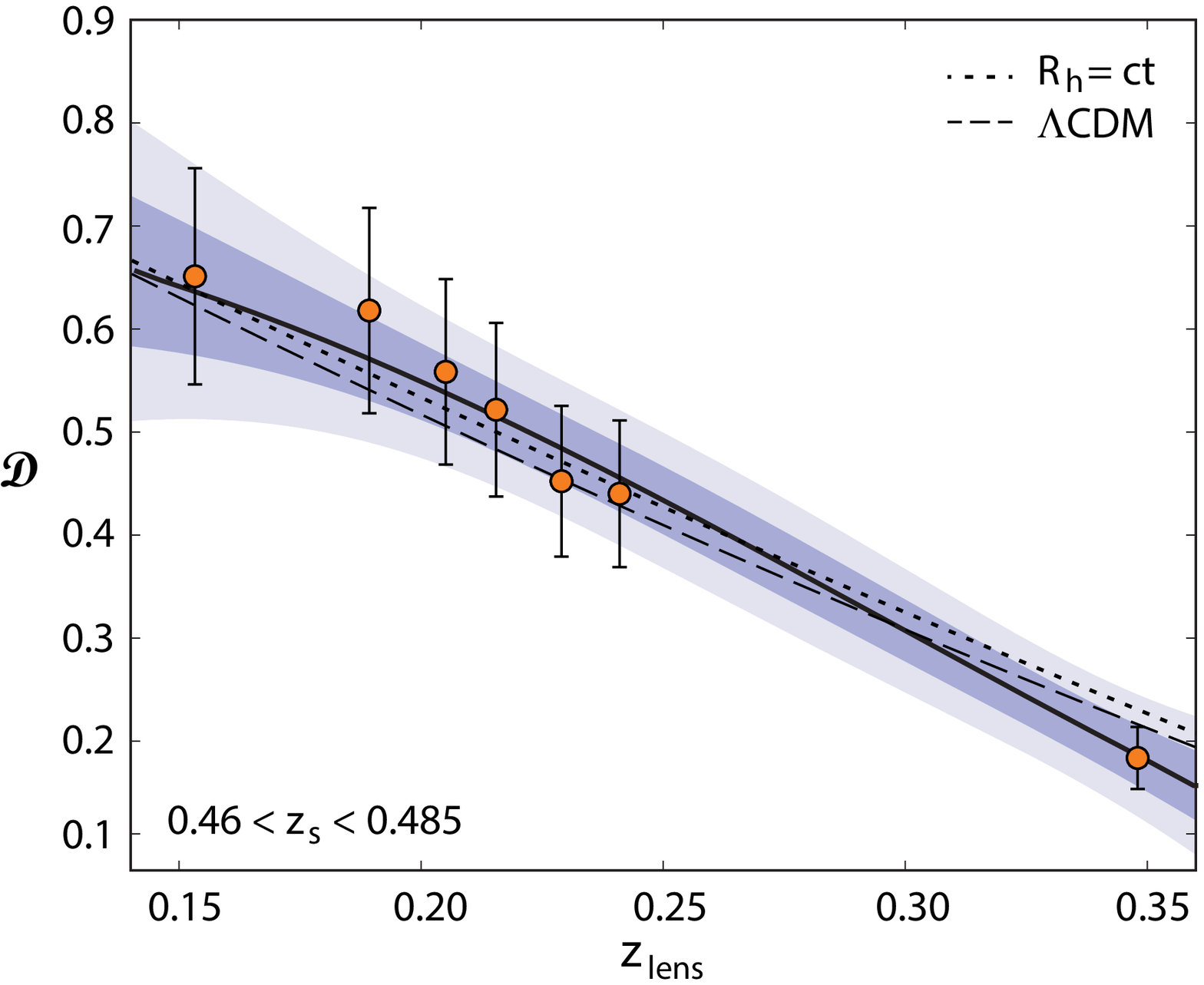}\hskip 0.1in 
\includegraphics[width=0.47\linewidth]{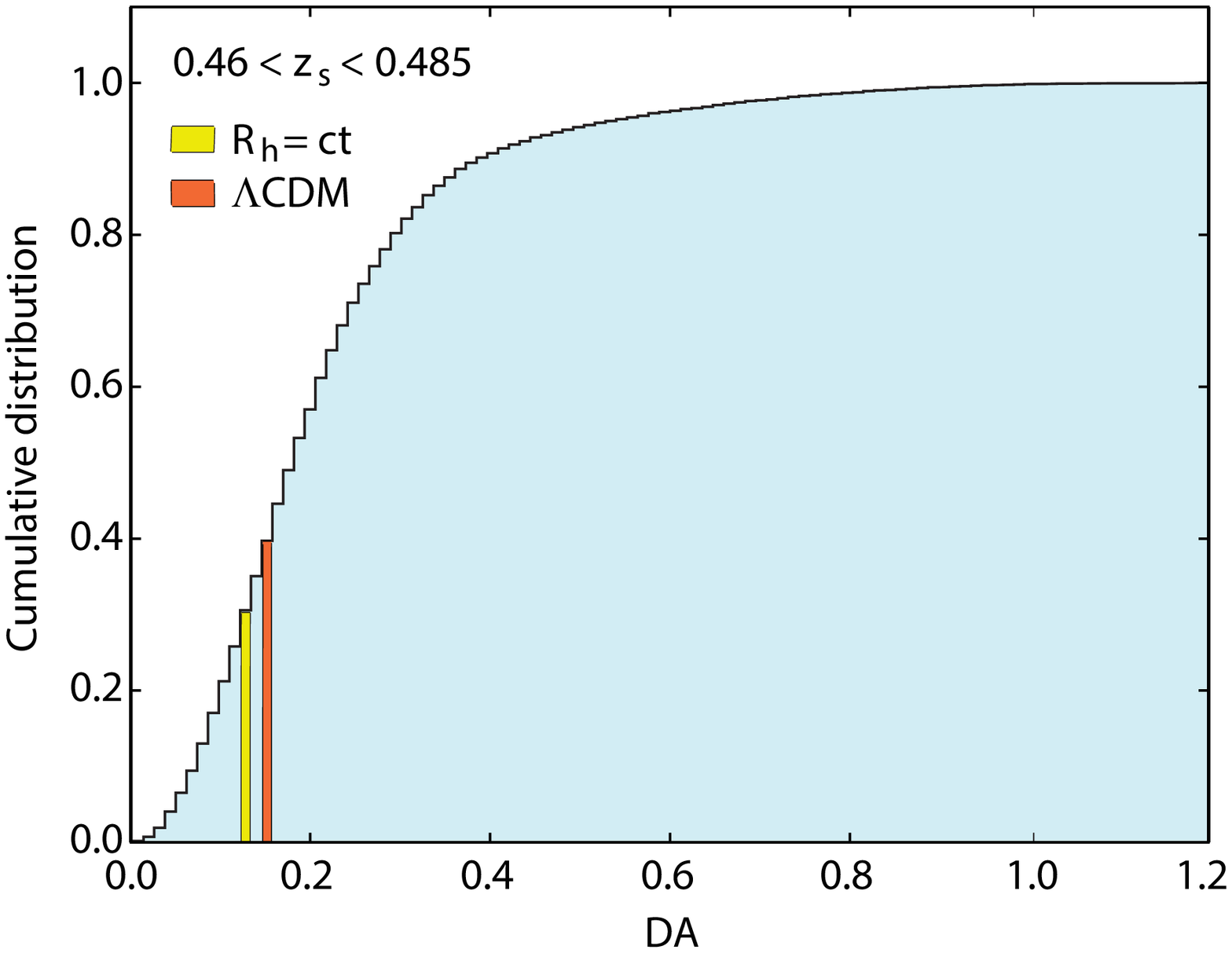} \\
\end{tabular}
\end{center}
\caption{{\it Left Panels} (2a, 2c, 2e, 2g, 2i): The solid curve in each plot indicates 
the reconstructed $\mathcal{D}_{\rm obs}$ function using Gaussian processes,
for the source redsfhit ranges $(0.51,0.535)$, $(0.52,0.545)$,
$(0.45,0.475)$, $(0.5,0.525)$, and $(0.46,0.485)$. The
dotted curve indicates the predicted $\mathcal{D}$ in the $R_{\rm h}=ct$
universe and the dashed curve indicates the corresponding $\mathcal{D}$
in $\Lambda$CDM. In each of these figures, dark blue represents the
$1\sigma$ confidence region, and light blue is $2\sigma$. {\it Right Panels}
(2b, 2d, 2f, 2h, 2j): The corresponding cumulative probability distributions.}
\end{figure}

\begin{table*}
\center
  \centerline{{\bf Table 6.} Model Comparison using Strong Gravitational Lenses with Gaussian Processes}\vskip 0.1in
\footnotesize
  \begin{tabular}{lcccc}
                \hline
                \hline
                && \\
Source Redshift Range & \qquad Number of Lenses & \qquad $R_{\rm h}=ct$ & \qquad $\Lambda$CDM & \qquad Figures \\
&&\qquad Probability&\qquad Probability& \\
\hline
&& \\
$0.51-0.535$ &\qquad 9 &\qquad $94.23\%$ &\qquad  $65.82\%$ &\qquad 2a, 2b \\
$0.52-0.545$ &\qquad 9 &\qquad $93.59\%$ &\qquad   $85.48\%$ &\qquad 2c, 2d \\
$0.46-0.485$ &\qquad 7 &\qquad $69.43\%$ &\qquad  $60.29\%$ &\qquad 2e, 2f \\
$0.50-0.525$ &\qquad 6 &\qquad $86.96\%$ &\qquad $49.91\%$ &\qquad 2g, 2h \\
$0.45-0.475$ &\qquad 5 &\qquad $80.65\%$ &\qquad  $69.68\%$ &\qquad 2i, 2j \\
&& \\
\hline\hline
  \end{tabular}
\end{table*}

\section{Conclusions}
In this paper, we have introduced the GP reconstruction approach to
strong lensing studies, though clearly the available sample is still
not large enough for us to make full use of this method. As noted
earlier, one of the principal benefits of this technique is that
the function (in this case $\mathcal{D}_{\rm obs}$) representing the
data may be obtained without the assumption of any parametric form
associated with particular models. This allows one to test different
models against the actual $\mathcal{D}_{\rm obs}$, rather than
against each other's predictions, neither of which may be a good
representation of the measurements. In addition, GP provide
$1\sigma$ and $2\sigma$ confidence regions for the reconstructed
functions more in line with the population as a whole, rather than
individual data points, greatly restricting the ability of
`incorrect' models to adequately fit the observations due to
otherwise large measurement errors.

This is reflected in the probabilities quoted in Table~6 for the
two models we have examined here. Unlike previous model comparisons
based on the use of parametric fits to the strong-lensing data,
we now find that $R_{\rm h}=ct$ is favoured over $\Lambda$CDM
with consistently higher likelihoods in all 5 source redshift
shells we have assembled for this work. Though these statistics
are still quite limited, it is nonetheless telling that the
differentiation between models improves as the number of sources
within each shell increases. Also, at least for $R_{\rm h}=ct$,
the probability of its predictions matching the GP reconstructed
functions generally increases as the size of the lens sample grows.
The outcome of this work underscores the importance of using
unbiased data and sound statistical methods when comparing different
cosmological models. As a counterexample, consider the use of
$H(z)$ measurements based on BAO observations instead of
cosmic chronometers \cite{28}, constituting an unwitting use
of model-dependent measurements to test competing models.
Such an approach ignores the significant limitations in all but the
three most recent BAO measurements \cite{58,59} for this type
of work. Previous applications of the galaxy two-point correlation
function to measure the BAO scale were contaminated with 
redshift distortions associated with internal gravitational
effects \cite{59}. To illustrate the significance of these
limitations, and the impact of the biased BAO measurements of
$H(z)$, note how the model favoured by the data switches from
$\Lambda$CDM to $R_{\rm h}=ct$ when only the unbiased measurements
are used \cite{61}.

A second counterexample is provided by the merger of disparate
sub-samples of Type Ia SNe to improve the statistical analysis. We
have already published an in-depth explanation of the perils associated
with the blending of data with different systematics for the purpose
of model selection \cite{62}, but let us nonetheless consider a brief
synopsis here.
The Union2.1 catalog \cite{63,64} includes $\approx580$ SN detections,
though each sub-sample has its own systematic and intrinsic
uncertainties. The conventional approach minimizes an overall~$\chi^2$,
while each sub-sample is assigned an intrinsic dispersion to ensure
that $\chi^2_{\rm dof}=1$ \cite{28,29}. Instead, the statistically
correct approach would estimate the unknown intrinsic dispersions
simultaneously with all other parameters \cite{62,65}.
Quite tellingly, the outcome of the model selection is reversed
when one switches from the improper statistical approach to the
correct one. To emphasize how critical this reversal is in the
case of $\Lambda$CDM, one simply needs to compare the outcome
of using a merged super-sample with that produced with a
large, single homogeneous sample, such as the Supernova Legacy
Survey Sample \cite{66}.

Within this context, we highlight the fact that the features
of the GP reconstruction approach in the study of
strong lenses are promising because, in spite of the fact that
the use of these systems to measure cosmological parameters has been
with us for over a decade (see, e.g., refs.~\cite{4,5,6,67}), the
results of this effort have thus far been less precise than those
of other kinds of observations. For the large part, these earlier
studies were based on the use of parametric fits to the data, but
it is quite evident (e.g., from figures~1 and 2 in ref.~\cite{7})
that the scatter in $\mathcal{D}_{\rm obs}$ about the theoretical
curves generally increases significantly as $D_A(z_l,z_s) \rightarrow 
D_A(0,z_s)$. That is, measuring $\mathcal{D}$ incurs a progressively
greater error as the distance to the gravitational lens becomes a
smaller fraction of the distance to the quasar source. This has
to do with the fact that $\theta_{\rm E}$ changes less for large
values of $z_s/z_l$ so, for a fixed error in the Einstein angle,
the measurement of $\mathcal{D}_{\rm obs}$ becomes less precise.
As we have demonstrated in this paper, the analysis of strong
lensing systems based on a GP reconstruction of
$\mathcal{D}_{\rm obs}$ improves our ability to distinguish
between different models, albeit by a modest amount given the
current sample.

Upcoming survey projects, such as the Dark Energy Survey (DES;
\cite{68}), the Large Synoptic Survey Telescope (LSST;
\cite{69}), the Joint Dark Energy Mission (JDEM; \cite{70}),
and the Square Kilometer Array (SKA; e.g., ref.~\cite{71}),
are expected to greatly grow the size of the lens sample.
The ability of GP reconstruction methods to differentiate between
models will increase in tandem with this growth. Several sources
of uncertainty still remain, however, including the actual mass
distribution within the lens. And since such errors appear to be
more restricting for lens systems with large values of $z_s/z_l$,
a priority for future work should be the identification of strong
lenses with small angular diameter distances between the source
and lens relative to the distance between the lens and observer.
 
{\acknowledgement
FM is grateful to the Instituto de Astrof\'isica
de Canarias in Tenerife and to Purple Mountain Observatory in Nanjing, China
for their hospitality while part of this research was carried out. FM is also
grateful for partial support to the Chinese Academy of Sciences Visiting
Professorships for Senior International Scientists under grant 2012T1J0011,
and to the Chinese State Administration of Foreign Experts Affairs under
grant GDJ20120491013.
\endacknowledgement}

%
%


\begin{thebibliography}{99}

\bibitem{1} M. Bartelmann \& P. Schneider, A\&A {\bf 345} (1999) 17
\bibitem{2} A. Refreiger, ARA\&A {\bf 41} (2003) 645
\bibitem{3} T. Futamase \& S. Yoshida, S., PThPh {\bf 105} (2001) 887
\bibitem{4} T. Treu, L.V.E. Koopmans, A. S. Bolton, S. Burles \& L. A. Moustakas, ApJ {\bf 640} (2006) 662
\bibitem{5} C. Grillo, M. Lombardi \& G. Bertin, A\&A {\bf 477} (2008) 397
\bibitem{6} M. Biesiada, A. Pi\'{o}rkowska \& B. Malec, B., MNRAS {\bf 406} (2010) 1055
\bibitem{7} F. Melia, J.-J. Wei \& X.-F. Wu, AJ {\bf 149} (2015) 2
\bibitem{8} S. Cao et al., ApJ {\bf 806} (2015) 185
\bibitem{9} Y. Shu et al., ApJ {\bf 851} (2017) 48
\bibitem{10} A. Bolton, S. M. Burles, L.V.E. Koopmans, T. Treu \& L. M. Moustakas, ApJ {\bf 638} (2006) 703
\bibitem{11} L.V.E. Koopmans, T. Treu, A. S. Bolton, S. Burles \& L. A. Moustakas, ApJ {\bf 649} (2006) 599
\bibitem{12} A. S. Bolton, S. M. Burles, L.V.E. Koopmans et al., ApJ {\bf 682} (2008) 964
\bibitem{13} E. R. Newton, P. J. Marshall \& T. Treu, (SLACS Collaboration), ApJ {\bf 734} (2011) 104
\bibitem{14} A.~G. Riess et al., AJ {\bf 116} (1998) 1009
\bibitem{15} S. Perlmutter et al., Nature {\bf 391} (1998) 51
\bibitem{16} F. Melia, MNRAS {\bf 382} (2007) 1917
\bibitem{17} F. Melia \& A. Shevchuk, MNRAS {\bf 419} (2012) 2579
\bibitem{18} F. Melia, MNRAS {\bf 464} (2017) 1966
\bibitem{19} F. Melia, ApJ {\bf 764} (2013) 72
\bibitem{20} J.-J. Wei, X. Wu \& F. Melia, ApJ {\bf 772} (2013) 43
\bibitem{21} J.-J. Wei, X.-F. Wu, F. Melia \& R. S. Maier, AJ {\bf 149} (2015) 102
\bibitem{22} F. Melia, J.-J. Wei, R. S. Maier \& X. Wu, AJ, submitted (2017)
\bibitem{23} F. Melia R. S. Maier, MNRAS {\bf 432} (2013) 2669
\bibitem{24} P. van Oirschot, J. Kwan \& G. F. Lewis, MNRAS {\bf 404} (2010) 1633
\bibitem{25} M. Bilicki \& M. Seikel, MNRAS {\bf 425} (2012) 1664
\bibitem{26} G. F. Lewis \& P. van Oirschot, MNRAS Letters {\bf 423} (2012) 26
\bibitem{27} G. F. Lewis, MNRAS {\bf 432} (2013) 2324
\bibitem{28} D. L. Shafer, Phys. Rev. D {\bf 91} (2015) 103516
\bibitem{29} D. Rubin \& B. Hayden B., ApJL {\bf 833} (2016) id. L30
\bibitem{30} O. Bikwa, F. Melia \& A.S.H. Shevchuk, MNRAS {\bf 421} (2012) 3356
\bibitem{31} F. Melia, JCAP {\bf 09} (2012) 029
\bibitem{32} F. Melia, Astrop. \& Sp. Sc. {\bf 356} (2015) 393
\bibitem{33} F. Melia, MNRAS {\bf 446} (2015) 1191
\bibitem{34} K. U. Ratnatunga, R. E. Griffiths \& E. J. Ostrander, AJ {\bf 117} (1999) 2010
\bibitem{35} P. Schneider, J. Ehlers \& E. E. Falco, ``Gravitational Lenses" (Berlin:Springer) (1992)
\bibitem{36} R. Kormann, P. Schneider \& M. Bartelmann, A\&A {\bf 284} (1994) 285
\bibitem{37} R. E. White \& D. S. Davis, BAAS {\bf 28} (1996) 1323
\bibitem{38} A. S. Bolton, S. M. Burles, L.V.E. Koopmans, T. Treu \& L. M. Moustakas, ApJL {\bf 624} (2005) L21
\bibitem{39} G. van de Ven, P. G. van Dokkum \& M. Franx, MNRAS {\bf 344} (2003) 924
\bibitem{40} S. Cao, Y. Pan, M. Biesiada, W. Godlowski \& Z.-H. Zhu, JCAP {\bf 3} (2012) 16
\bibitem{41} P. Young, J. E. Gunn, J. Kristian, J. B. Oke \& J. A. Westphal, ApJ {\bf 241} (1980) 507
\bibitem{42} J. Huchra, M. Gorenstein, S. Kent et al., AJ {\bf 90} (1985) 691
\bibitem{43} J. Lehar, G. I. Langston, A. Silber, C. R. Lawrence \& B. F. Burke, AJ {\bf 105} (1993) 847
\bibitem{44} C. D. Fassnacht, D. S. Womble, G. Neugebauer et al., ApJL {\bf 460} (1996) L103
\bibitem{45} J. L. Tonry, AJ {\bf 115} (1998) 1
\bibitem{46} L.V.E. Koopmans \& T. Treu, ApJ {\bf 568} (2002) 5
\bibitem{47} L.V.E. Koopmans \& T. Treu, ApJ {\bf 583} (2003) 606
\bibitem{48} T. Treu \& L.V.E. Koopmans, ApJ {\bf 611} (2004) 739
\bibitem{49} T. Treu \& L.V.E. Koopmans, MNRAS {\bf 343} (2003) 29
\bibitem{50} T. Treu \& L.V.E. Koopmans, ApJ {\bf 575} (2002) 87
\bibitem{51} M. Seikel, C. Clarkson \& M. Smith, M., JCAP {\bf 06} (2012) 036S
\bibitem{52} F. Melia \& M. Abdelqader, IJMP-D {\bf 18} (2009) 1889
\bibitem{53} F. Melia, Frontiers of Physics {\bf 11} (2016) 119801
\bibitem{54} F. Melia, Fronters of Physics {\bf 12} (2017) 129802
\bibitem{55} F. Melia \& M. Fatuzzo, MNRAS {\bf 456} (2016) 3422
\bibitem{56} Yennapureddy, M. K. \& Melia F, JCAP {\bf 11} (2017) 029Y
\bibitem{57} F.Melia \& Yennapureddy, M. K, JCAP {\bf 02} (2018) 034M
\bibitem{58} L. Anderson et al., MNRAS {\bf 441} (2014) 24
\bibitem{59} T. Delubac et al., A\&A {\bf 574} (2015) A59
\bibitem{60} M. L\'opez-Corredoira, ApJ {\bf 781} (2014) id. 96
\bibitem{61} F. Melia \& M. L\'opez-Corredoira, IJMP-D {\bf 26} (2017) 1750055
\bibitem{62} J.-J. Wei, X. Wu, F. Melia \& R. S. Maier, AJ {\bf 149} (2015) 102
\bibitem{63} M. R. Kowalski, ApJ {\bf 686} (2008) 749
\bibitem{64} N. Suzuki et al., ApJ {\bf 746} (2012) 85
\bibitem{65} A. G. Kim, PASP {\bf 123} (2011) 230
\bibitem{66} M. Betoule et al., A\&A {\bf 568} (2014) 22
\bibitem{67} M. Biesiada, B. Malec \& A. Pi\'orkowska, RAA {\bf 11} (2011) 641
\bibitem{68} J. Frieman \& Dark Energy Survey Collaboration, BAAS {\bf 36} (2004) 1462
\bibitem{69} A. Tyson \& R. Angel, in The New Era of Wide Field Astronomy, ASP
Conference Series, Vol. 232. Ed. R. Clowes, A. Adamson \& G. Bromage. San Francisco: ASP, p.347 (2001)
\bibitem{70} A. Tyson, in ASP Conference Series, Vol. 339, Observing Dark Energy,
Ed. S. C. Wolff \& T. R. Lauer, p. 95 (2005)
\bibitem{71} J. McKean, N. Jackson, S. Vegetti, M. Rybak, S. Serjeant,
L.V.E. Koopmans, R. B. Metcalf, C. Fassnacht, P. J. Marshall \& M. Pandey-Pommier,
in Proceedings of Advancing Astrophysics with the Square Kilometre Array (AASKA14). 9-13 June, 2014. Giardini
Naxos, Italy (2015)

\end{thebibliography}
\end{document}